\documentclass[epj]{svjour}
\usepackage{graphics}

\begin{document}

\title{Measurement of the isospin-filtering \boldmath
  $\mathsf{dd\to{}^4He\,K^+K^-}$ reaction at $\mathsf{Q=39}$~MeV}
\titlerunning{The isospin-filtering $dd\to{}^4\mathrm{He}\, K^+K^-$
  reaction}
\authorrunning{X.~Yuan \textit{et al.}}

\author{
  X.~Yuan\inst{1,2} \and
  M.~B\"uscher\inst{2} \and
  D.~Chiladze\inst{2,3}
  S.~Dymov\inst{4,5} \and
  A.~Dzyuba\inst{2,6} \and
  V.Yu.~Grishina\inst{7} \and
  C.~Hanhart\inst{2,8} \and
  M.~Hartmann\inst{2} \and
  A.~Kacharava\inst{2} \and
  A.~Khoukaz\inst{9} \and
  L.A.~Kondratyuk\inst{10} \and
  V.~Koptev\inst{6} \and
  P.~Kulessa\inst{11} \and
  M.~Mielke\inst{9} \and
  S.~Mikirtychiants\inst{2,6} \and
  M.~Nekipelov\inst{2} \and
  M.~Papenbrock\inst{9} \and
  K.~Pysz\inst{11} \and
  V.~Serdyuk\inst{2,4} \and
  H.~Str\"{o}her\inst{2} \and
  C.~Wilkin\inst{12}\thanks{e-mail: \texttt{cw@hep.ucl.ac.uk}} \and
  H.~Xu\inst{1}
}

\institute{
  Institute of Modern Physics, Chinese Academy of Sciences,
  Lanzhou 730000, P.R.~China \and
  Institut f\"ur Kernphysik and J\"ulich Centre for Hadron Physics,
  Forschungszentrum J\"ulich, D-52425 J\"ulich, Germany \and
  High Energy Physics Institute, Tbilisi State University, 
  GE-0186 Tbilisi, Georgia \and
  Laboratory of Nuclear Problems, Joint Institute for Nuclear Research, 
  RU-141980 Dubna, Russia \and
  Physikalisches Institut II, Universit\"{a}t Erlangen-N\"{u}rnberg, D-91058,
  Erlangen, Germany \and
  Petersburg Nuclear Physics Institute, RU-188300 Gatchina, Russia \and
  Institute for Nuclear Research, RU-117312 Moscow, Russia \and
  Institute for Advanced Simulations,
  Forschungszentrum J\"ulich, D-52425 J\"ulich, Germany \and
  Institut f\"ur Kernphysik, Universit\"at M\"unster, 
  D-48149 M\"unster, Germany \and
  Institute of Theoretical and Experimental Physics,
  RU-117218 Moscow, Russia \and
  H.~Niewodnicza\'{n}ski Institute of Nuclear Physics PAN,
  PL-31342 Krak\'{o}w, Poland \and
  Physics and Astronomy Department, UCL, London, WC1E 6BT, UK
}

\date{\today}

\abstract{The total cross section for the $dd\to{}^4\mathrm{He}\,
  K^+K^-$ reaction has been measured at a beam momentum of
  3.7\,GeV/$c$, corresponding to an excess energy of 39\,MeV, which is
  the maximum possible at the Cooler Synchrotron COSY-J\"{u}lich. A
  deuterium cluster-jet target and the ANKE forward magnetic
  spectrometer, placed inside the storage ring, have been employed in
  this investigation. We find a total cross section of
  $\sigma_{\mathrm{tot}} < 14\,\textrm{pb}$, which brings into question
  the viability of investigating the $dd\to{}^4\mathrm{He}\,a_0(980)$
  reaction as a means of studying isospin violation.
\PACS{{14.40.Cs}{Other mesons with $S=C=0$, mass $< 2.5\,$GeV}\and %
{25.45.-z}{2H-induced reactions}
 }
} 

\maketitle

%
%
\section{Introduction}
\label{intro}

Since the deuteron and the $\alpha$-particle both have isospin $I=0$,
the $dd\to{}^4\mathrm{He}\,X$ reaction provides an ``isospin filter''
that favours $I=0$ states $X$. The first definitive measurement of
the $dd\to{}^4\mathrm{He}\,\pi^0$ cross section near
threshold~\cite{stephenson} has therefore provided clear evidence for
isospin violation in the form of charge symmetry breaking (CSB).

Within the standard model there are two dominating sources of isospin
violation, namely the electromagnetic interaction and the differences
in the masses of the quarks~\cite{miller}. One way in which the quarks
may play a role is through inducing a mixing between $I=0$ and $I=1$
mesons, such as that between the $\eta$ and the $\pi^0$. In this sense
the observed mesons are not pure isospin eigenstates and one
contribution to the CSB reaction would be through the virtual
production of $\eta$ meson which mixes to emerge as a $\pi^0$.
However, as a result of the relatively large mass of the strange
quark, which leads to a significant mass splitting of $\pi^0$ and
$\eta$, the effect of the $\eta/\pi^0$ mixing is not expected to be
dominant in isospin-violating pion
production~\cite{vanKolck:2000ip}. For a discussion and exploratory
calculations for the $dd\to{}^4\mathrm{He}\,\pi^0$  reaction, see
Refs.~\cite{Gardestig:2004hs,Theory}.

On the other hand the $a_0(980)$ and $f_0(980)$ scalar mes\-ons, which
have $I=1$ and $I=0$ respectively, are rather narrow but overlapping
resonances.  This alone already enhances the effect of meson mixing in
the final state compared to isospin violation in the production
operator~\cite{Grishina,Hanhart:2003pg}. In addition, since both
mesons couple strongly to $K\bar{K}$~\cite{PDG}, a particularly large
contribution coming from kaon loops~\cite{Achasov:1979xc,CH}
significantly enhances the mixing amplitude. Therefore one might hope
to get a direct experimental access to the $a_0/f_0$ mixing amplitude.

The main decay channel of the $I=1$ meson is through $a_0(980)\to
\eta\pi$~\cite{PDG} and so a promising measurement of CSB might be
through the study of $dd \to {}^4\mathrm{He}\,f_0(980) \to
{}^4\mathrm{He}\,a_0(980) \to {}^4\mathrm{He}\,
\eta\pi^0$~\cite{WASA}, with an isospin-allowed $f_0(980)$ production
in the first step. Whether this promise can be fulfilled or not would
depend to a large extent on the production cross section for the
$f_0(980)$ meson in this channel. It is the primary aim of the
current work to provide experimental data on this through a
measurement of the $dd\to{}^4\mathrm{He}f_0\to{}^4\mathrm{He}\,K^+K^-$ cross section.

The conditions of the experiment, which was carried out using the
ANKE spectrometer at COSY-J\"{u}lich, are described in
sect.~\ref{experiment}. The strict conditions on the particle
selection necessary to identify the desired reaction against a much
larger background are here explained in some detail. Two
methodologies that remove almost all the background are discussed
but, in view of the consequent low acceptances, they could only
provide upper limits on the production rate. By relaxing the
criteria, it proved possible to extract events with a reasonable
background and hence to evaluate the total cross section which is reported in
sect.~\ref{sec:acceptance}. The smallness of the resulting value
suggests that the $dd\to{}^4\mathrm{He}\,\eta\pi^0$ reaction might in
fact not be a useful way of investigating isospin violation and this
is reflected in our conclusions of sect.~\ref{Conclusions}.

%
%
\section{Identification of \boldmath $\mathsf{dd\to{}^4He\,K^+K^-}$
events with ANKE}%
\setcounter{equation}{0}%
\label{experiment}
%
%
\subsection{Experimental setup}
The COSY COoler SYnchrotron of the Forschungszentrum J\"ulich can
accelerate and store deuterons with momenta of up to
3.7\,GeV/\textit{c}~\cite{COSY}, which corresponds to an excess energy
of $Q=39$\,MeV for the $dd\to{}^4\mathrm{He}\, K^+K^-$ reaction. The
studies of this reaction were carried out at the ANKE
facility~\cite{anke1}, a magnetic spectrometer located in one of the
straight sections of COSY. ANKE has three dipole magnets, D1 -- D3.
D1 deflects the circulating COSY beam onto a target in front of D2,
and D3 bends it back into the nominal orbit. The C-shaped spectrometer
dipole D2 separates forward-going reaction products from the COSY beam
and allows one to determine their emission angles and momenta.

The deuterium cluster-jet target~\cite{target} used with ANKE had an
areal density of a few times $10^{14}$\,cm$^{-2}$ which, combined
with a typical deuteron beam intensity of
$(3-6)\times$10$^{10}$\,s$^{-1}$, yields an average luminosity of
$L=[2.6\pm 0.1\mathrm{(stat)} \pm 0.8 \mathrm{(syst)} \pm
\mathrm{0.3(syst)}]\times 10^{31}\, \mathrm{s}^{-1}
\mathrm{cm}^{-2}$~\cite{lumi}.  This gave an integrated luminosity of
$L_{\mathrm{int}} = 35$\,pb$^{-1}$ over the course of the experiment.
The two systematic errors are due mainly to the uncertainty in the
quasi-elastic $dd\to dX$ cross section used for the normalisation and
to a possible error in the determination of the angle-momentum
acceptance for that process in ANKE.

%
%
\subsection{Particle selection criteria}
In order to isolate the $dd\to{}^4\mathrm{He}\, K^+K^-$ events, two
well-identified final-state particles, \emph{viz} the $K^+$ and
$^4$He, could be detected in coincidence, with the remaining $K^-$
meson being recognised through the missing-mass in the reaction.
Alternatively, all three final particles could be measured with rather
looser criteria. Both approaches have been implemented in this work.

\begin{figure}
  \resizebox{0.95\linewidth}{!}{%
    \includegraphics{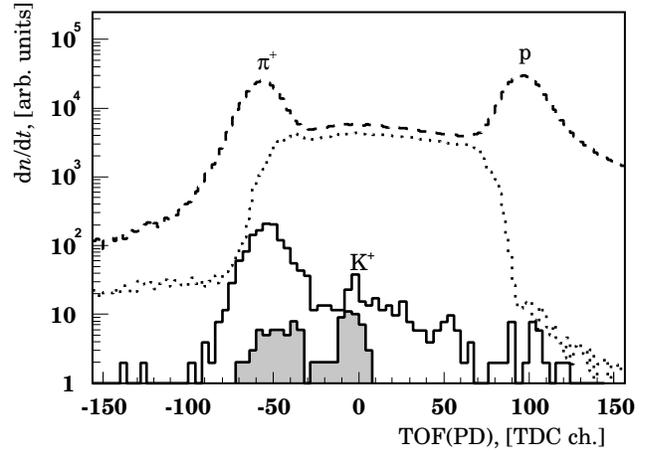}}
  \caption{Time-of-flight spectrum in the ANKE positive detector (PD)
    for increasingly stringent on-line trigger settings and
    $K^+$ selection criteria. Dashed line: open trigger run; dotted
    line: on-line TOF trigger for the detection of $K^+$ mesons; solid
    line: same as dotted plus a $K^+$-cut on the time difference
    between the PD and forward detector (FD); shaded histogram: same
    as solid plus the delayed veto criterion in the ANKE range
    telescopes. In all cases a valid track (reconstructed from the
    MWPC information) in the PD and FD was required and an
    energy-loss cut on deuterons (and heavier particles) in the FD
    (cf.~Fig.~\ref{fig:alpha_id}) has been applied. The spectra are
    normalised to the relative luminosities and dead times.}
  \label{fig:K+_id}
\end{figure}

Positively charged kaons can be identified in the positive side
detection system (PD)~\cite{anke1,anke2} of ANKE by a time-of-flight
(TOF) measurement, as illustrated in Fig.~\ref{fig:K+_id}. The TOF
start counters, consisting of one layer of 23 scintillation counters,
are mounted next to the large exit window of the vacuum chamber in
D2. Positive kaons from the $dd\to{}^4\mathrm{He}\,K^+K^-$ reaction
with momenta between 390 and 625\,MeV/$c$ stop in the range
telescopes located along the focal surface of D2. These telescopes
serve as the TOF stop counters and provide further, extremely
effective, kaon-\emph{versus}-background discrimination through the
measurement of delayed signals from the $K^+$ decay products, $\mu^+$
and $\pi^+$~\cite{anke2}. A drawback of this method is its relatively
low $K^+$-detection efficiency of only about 15\%.  Two multi-wire
proportional chambers (MWPCs) positioned between the TOF start and
stop counters allow one to deduce the ejectile momenta and to
suppress the background from secondary scattering.

A rather strict $K^+$ TOF criterion was already included in the
on-line trigger that suppresses $\pi^+$ mesons and protons. In order
to collect data for the calibration of the $^3$He and $^4$He energy
losses, several runs were taken with an open TOF trigger
(cf.~Fig.~\ref{fig:K+_id}).

High momentum particles, such as protons, deuterons and He nuclei,
that are produced in coincidence with the $K^+$ mesons, as well as
scattered deuterons and protons from the breakup of the beam
deuterons, were detected in the forward-detector (FD)~\cite{lumi,fwd}.
This contains three layers of scintillation counters for TOF and
energy-loss ($\Delta E$) measurements. In addition, there are three
MWPCs that are used for momentum reconstruction and background
suppression.

Negatively charged particles can be measured in the negative-particle
detection system (ND)~\cite{anke3}, comprising scintillators for TOF
measurements and two MWPCs for tracking. The $K^-$ are separated from
the $\pi^-$ background by TOF criteria and, in addition, a
reconstructed track from the ND MWPCs is required.

In addition to the above mentioned $K^+$ TOF criterion, the energy
loss in the FD was already used at the trigger level. The $\Delta E$
thresholds were set such that events with $^4$He particles as well as
deuterons were retained for further analysis. The latter were used
for a fine-tuning of the $K^+$ selection criteria, since $dK^+$ pairs
are produced at a much higher rate than the $^4\mathrm{He}\,K^+$
events of interest.

Events with a $K^+$ meson in the PD and a deuteron in the FD can be
identified from the relative detection times in these counters. This
time difference, $\Delta t_{\mathrm{FD,PD}}(d,K^+)$, can be determined
either directly from the calibrated TDC data or reconstructed from the
particle trajectories and momenta, as deduced from the MWPC track
information. For particles arising from the same reaction vertex,
these values are correlated~\cite{anke3}. The solid line in
Fig.~\ref{fig:K+_id} shows the PD TOF spectrum after the application
of the $\Delta t_{\mathrm{FD,PD}}(d,K^+)$ cut. A clear $K^+$ peak from
the $dK^+$ events is seen, though a substantial background from
secondary particles remains.

Figure~\ref{fig:K+_id} shows that, when the delayed-veto cut for the
$K^+$ mesons is applied in addition to the $\Delta
t_{\mathrm{FD,PD}}(d,K^+)$ criterion, a background-free identification
of $dK^+$ correlations is possible.

\begin{figure}
  \resizebox{0.95\linewidth}{!}{%
    \includegraphics{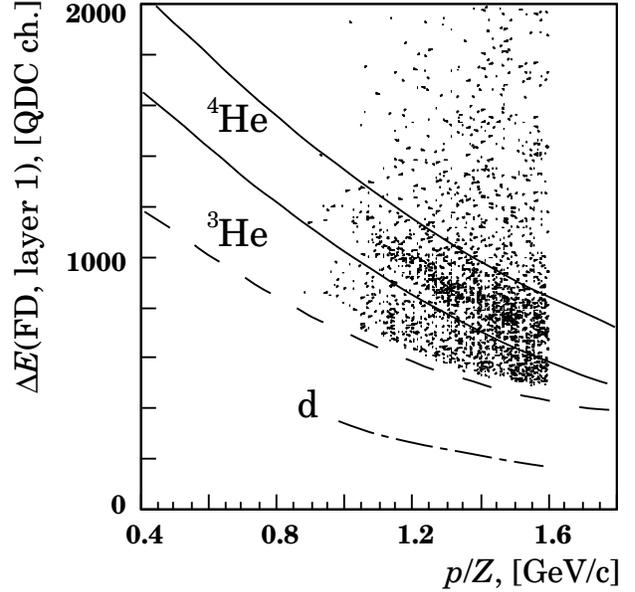} }
  \caption{Identification of $^4$He particles (selected in coincidence
    with $\pi^+$ mesons) using energy losses in the FD scintillation
    counters (shown here for scintillator layer 1). The solid lines
    show the boundaries of the $\Delta E$ cut used for $^4$He
    selection. The dashed and dash-dotted lines indicate the positions
    of the centres of the $^3$He and deuteron bands, respectively.
    For a better visibility of the $^4$He energy-loss band, only those
    events are shown that fulfil the following criteria: \textit{i})
    an energy loss above the $^3$He band in layer 1; \textit{ii}) a
    valid $\Delta t_{\mathrm{FD,PD}}$ for $^4$He$\,\pi^+$ events for
    layers 1 and 2; \textit{iii}) a cut on the energy loss of layer 2
    for the $^4$He candidates.}
  \label{fig:alpha_id}
\end{figure}

Since the $^4$He particles have both twice the charge and almost
twice the mass of the deuterons, their time correlations with the
kaons are indistinguishable. Therefore, some of the nominally $K^+d$
events in Fig.~\ref{fig:K+_id}, selected on the basis of the $\Delta
t_{\mathrm{FD,PD}}(d,K^+)$ cut, may stem from the
$dd\to{}^4\mathrm{He}\,K^+ X$ reaction. In order therefore to
discriminate between $^4$He and deuterons, a rigidity-dependent
$\Delta E$ cut has been applied in all FD scintillators, as shown in
Fig.~\ref{fig:alpha_id}. The exact locations of the $\Delta
E$-\emph{versus}-$\vert\vec{p}_{FD}\vert/Z$ bands have been
determined from events with a coincident $\pi^+$ meson in the PD
(open TOF trigger runs). Their positions are in good agreement with
those obtained from Monte-Carlo simulations.

%
%
\subsection{Event selection with maximum background suppression}
\label{sec:maxbg}

For the identification of the $dd\to{}^4\mathrm{He}\,K^+K^-$ events
the following $K^+$ and $^4$He selection criteria are jointly used,
as described above:
\begin{tabbing}
\emph{i})\hspace{3mm}\=TOF for $K^+$ mesons in the PD;\\ %
\emph{ii})\>time difference between particle detection in the PD\\ %
           \>and FD, $\Delta t_{\mathrm{FD,PD}}(^4\mathrm{He},K^+)$;\\ %
\emph{iii})\> $\Delta E$ \emph{versus} $\vert\vec{p}_{FD}\vert/Z$ in
the FD;\\ %
\emph{iv})\>delayed-veto signal in the range telescopes.
\end{tabbing}
These combined criteria offer the maximum possible background
suppression at ANKE, though at the expense of having the low
efficiency ($\sim 15$\%) of the delayed-veto criterion. However, it
turned out that no events survive these criteria and this procedure
only leads to an upper limit on the production cross section.

Another possibility to provide an efficient background suppression
with comparable detection efficiency is to abandon the delayed-veto
criterion and to search directly for $K^-$ mesons in correlation with
the detected $^4\mathrm{He}\,K^+$ pairs. The additional demands
imposed to ensure that any signal comes from a $K^-$ meson reduces
the total acceptance by a factor $\sim 5$.
One event was found by this method and this is clearly consistent
with the zero events obtained with the delayed-veto procedure.

The above considerations show that only upper limits on the total
$dd\to{}^4\mathrm{He}\,K^+K^-$ production cross section can be
deduced by using ``background-free'' methods and that some of the
criteria must be relaxed in order to obtain a signal.
%
%
\subsection{Event selection with relaxed cuts}
\label{sec:relbg}

\begin{figure*}
  \resizebox{\linewidth}{!}{%
   \includegraphics{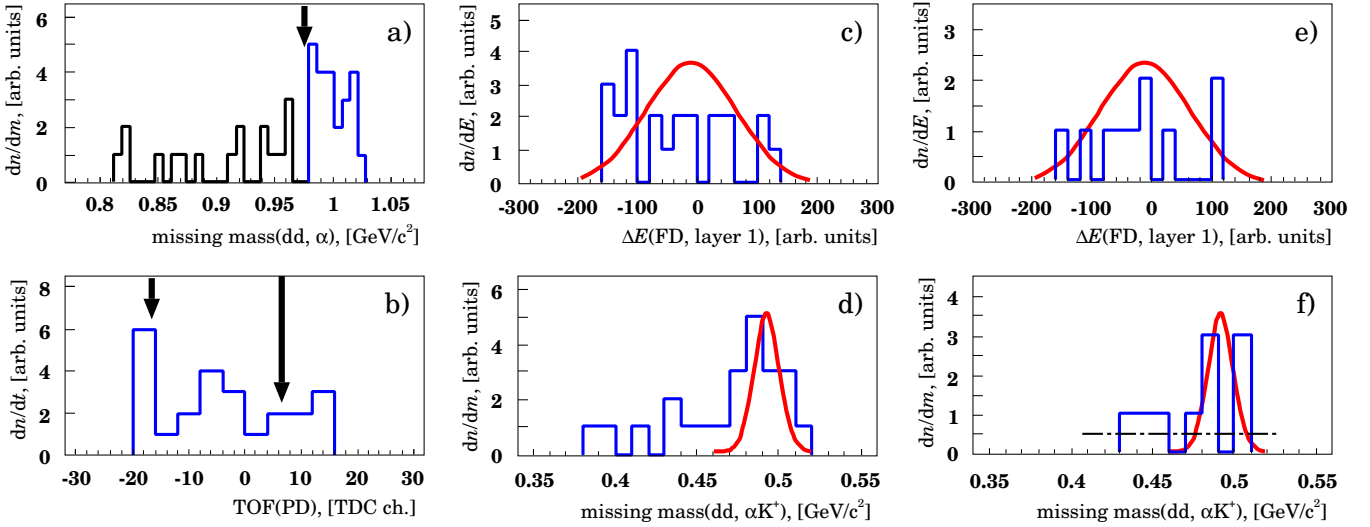}}
  \caption{a)--d) Events remaining after the application of the
    $K^+$ TOF cut (Fig.~\ref{fig:K+_id}), the time correlation between
    the PD and FD, and the $^4$He energy-loss criterion
    (Fig.~\ref{fig:alpha_id}). For panels b)--d) an additional ($dd$,
    $^4$He) missing-mass cut $m>980$\,MeV/$c^2$ has been used, as
    indicated in panel a).  Panels e) and f) show the residual events
    after the use of the stronger TOF cut indicated in panel b) by the
    arrows. Panels c) and e) show the difference between the measured
    and expected $^4$He energy loss.  The curves represent the shapes
    of the expected signals, as derived from Monte-Carlo simulations.}
  \label{fig:final}
\end{figure*}

Figure~\ref{fig:final}a) shows the $(dd,{}^4\mathrm{He})$ missing-mass
distribution for the events remaining after imposing the following
criteria:
\begin{tabbing}
\emph{i})\hspace{3mm}\=TOF for $K^+$ mesons in the PD;\\ %
\emph{ii})\>time difference between particle detection in the PD\\
\>and FD, $\Delta t_{\mathrm{FD,PD}}(^4\mathrm{He},K^+)$;\\ %
\emph{iii})\>$\Delta E$ \emph{versus}$\vert\vec{p}_{FD}\vert/Z$ in
the FD.%
\end{tabbing}

Figures~\ref{fig:final}b)--d) show the remaining events after
additionally demanding that the $(dd,^4\mathrm{He})$ missing mass must
be above the $K^+K^-$ threshold:
\begin{tabbing}
  \emph{iv$^{\,\prime}$}) \hspace{3mm}\=  $mm(dd,^4\mathrm{He})> m(K^+K^-)$  
  (arrow in Fig.~\ref{fig:final}a).%
\end{tabbing}

Compared to the ``background-free'' selection from the previous
section, we do not impose the delayed-veto criterion and use the
missing-mass cut on the $^4$He instead.

In order to suppress the residual background, an additional stronger
cut on the TOF between the start and stop counters in the PD has been
applied to select the $K^+$ mesons (arrows in Fig.~\ref{fig:final}b)
with efficiency $\varepsilon_{\mathrm{TOF(K^+)}}=90$\%. The results
are shown in Figs.~\ref{fig:final}e) and \ref{fig:final}f). 
Virtually identical distributions are obtained for a narrower $\Delta
t_{\mathrm{FD,PD}}(^4\mathrm{He},K^+)$ time gate.

Under the assumption that the background distribution for the $^4$He
missing-mass is flat and is at a level of 0.5 events per bin
(Fig.~\ref{fig:final}f), we find evidence for about 5 events coming
from the $dd\to{}^4\mathrm{He}\, K^+K^-$ reaction.  This number is
consistent with the upper limits derived using the ``background-free''
methods described in the previous section.

%
%
\section{Total production cross section}
\label{sec:acceptance} \setcounter{equation}{0}

In order to evaluate the total $dd\to{}^4\mathrm{He}\,K^+K^-$
production cross section, the geometrical acceptance of ANKE, the
detector resolutions and efficiencies, dead times, and kaon-decay
probabilities were taken into account in a Monte Carlo simulation,
written using the GEANT4 code~\cite{geant}. Two possibilities for the
input distributions have been considered, \emph{viz} three-body phase
space and a Flatt\'e $K^+K^-$ mass distribution~\cite{flatte}, where
the $f_0(980)$ parameters were those obtained by the BES
collaboration~\cite{bes}. The simulated results for these two
scenarios are basically identical and the average total acceptance
for the $^4\mathrm{He}\,K^+$ detection is estimated to be
$A_{^4\mathrm{He},K^+}=8.5$\%.

Taking the dead time into account, the efficiency of the ANKE data
acquisition system is $\varepsilon_{\mathrm{DAQ}}=70$\%, while the SD
and FD efficiencies are $\varepsilon_{\mathrm{SD}}=95$\% and
$\varepsilon_{\mathrm{FD}}=70$\%.

The two methods described in sect.~\ref{sec:maxbg} as being
``back\-ground-free'' are almost independent statistically and their
simultaneous measurement allows one to deduce an upper limit for the
$dd\to{}^4\mathrm{He}\,K^+K^-$ cross section at $Q = 39$\,MeV:
\begin{equation}
  \label{eq:upper_limit} \sigma_{\mathrm{tot}} < 14\,\textrm{pb}
\quad\quad (92\% \textrm{CL})
\end{equation}

On the other hand, the five events obtained after background
subtraction shown in Fig.~\ref{fig:final}f) lead to a total
production cross section of
\begin{equation}
  \label{eq:sigma_tot}
  \sigma_{\mathrm{tot}}=(5\pm 2_{\mathrm{stat}}\pm 1_{\mathrm{syst}})\,\textrm{pb}.
\end{equation}
with an additional overall uncertainty of about 50\% coming from the
luminosity determination. The two approaches give consistent results
and show that the $^{4}\textrm{He}\,K^+K^-$ production cross section
in deuteron-deu\-teron collisions near threshold is very low indeed.
%
%
\section{Discussion and outlook}
\label{Conclusions}\setcounter{equation}{0}

In order to investigate the production of the upper tail of the $f_0$
meson in deuteron-deuteron collisions, we have undertaken a
measurement of the $dd\to{}^4\mathrm{He}\,K^+K^-$ reaction at an
excess energy of 39\,MeV. The production cross is surprisingly low
such that the techniques which we used to eliminate the background
essentially eliminated the very weak signal as well. By relaxing the
selection criteria and subtracting the background, it was possible to
find five events that corresponded to a cross section of a mere 5\,pb.

One of the principal advantages of the reaction studied is that the
selection rules require the $K^+K^-$ pair to be in isospin $I=0$ but,
in view of the poor statistics,
there is no way of knowing whether the five events arise from the
production of scalar ($f_0$) pairs or vector ($\phi$) pairs. There is
evidence from the $pd\to{}^3\mathrm{He}\,K^+K^-$ reaction that even at
$Q=35$\,MeV the production of the $\phi$-meson represents a
significant fraction of the total cross section~\cite{MOMO}.

Since no theoretical models exist for $K^+K^-$ production in this
reaction, it is perhaps permissible to make a crude estimate based
upon the factorisation ansatz whereby
\begin{eqnarray}
  \nonumber \sigma_{\mathrm{tot}}(dd\to{}^4\mathrm{He}\,K^+K^-)= &&\\
  \label{ansatz} &&\hspace{-3.9cm}
  \frac{\sigma_{\mathrm{tot}}(dd\to{}^4\mathrm{He}\,\eta)}
  {\sigma_{\mathrm{tot}}(pd\to{}^3\mathrm{He}\,\eta)}
  \times\sigma_{\mathrm{tot}}(pd\to{}^3\mathrm{He}\,K^+K^-).
\end{eqnarray}

The production of the $\eta$ meson via $dd\to{}^4\mathrm{He}\,\eta$
has only been measured up to $Q=16.6$\,MeV, where the total cross
section is about 16\,nb~\cite{Budzanowski:2008qx}. Interpolating the
$pd\to{}^3\mathrm{He}\,\eta$ data~\cite{Mersmann,Jozef} to this energy
gives $\sigma_{\mathrm{tot}}(pd\to{}^3\mathrm{He}\,\eta)\approx
340$\,nb so that the simplistic approach of Eq.~(\ref{ansatz}) would
predict
\begin{eqnarray}
  \nonumber 
  \lefteqn{\sigma_{\mathrm{tot}}(dd\to{}^4\mathrm{He}\,K^+K^-)\approx}\\
  \label{predict}&&  0.05
  \times\sigma_{\mathrm{tot}}(pd\to{}^3\mathrm{He}\,K^+K^-).
\end{eqnarray}

Although the data are more sparse, a roughly similar ratio is obtained
if one uses the results on
$pd\to{}^3\mathrm{He}\,\omega$~\cite{Wurzinger} and
$dd\to{}^4\mathrm{He}\,\omega$~\cite{Banaigs}, though with larger
uncertainties.

It was shown in Ref.~\cite{MOMO} that the non-$\phi$ production cross
section varies as
\begin{equation}
  \label{nonphi}
  \sigma_{\mathrm{tot}}(pd\to{}^3\mathrm{He}\,K^+K^-)\approx
  5\,(Q/\textrm{MeV})^2\,\textrm{pb},
\end{equation}
which gives 7.5\,\textrm{nb} for our excess energy. These
considerations would lead us to expect a total cross section for
$dd\to{}^4\mathrm{He}\,K^+K^-$ of about 370\,pb, which is almost two
orders of magnitude larger than our measurement and does not even
allow for the contribution from $\phi$ production. This vast
discrepancy suggests that the dynamics governing the production of the
$\eta$ or $\omega$ meson and $K^+K^-$ pairs in these reactions are
intrinsically different and this may come about from the kaons being
produced at different vertices whereas a solitary meson must originate
from a single one.

Independent of the underlying dynamics, if one assumes that the five
events are products of the decay of the $f_0(980)$ resonance, which
has a branching ratio of $0.12\pm0.06$ into $K^+K^-$~\cite{bes}, the
total cross section for the production of this state through
$dd\to{}^4\mathrm{He}\,f_0(980)$ is only about 40\,pb at our energy.
This low value must bring into question any attempt to measure charge
symmetry breaking in the $dd\to{}^4\mathrm{He}\,\pi^0\eta$ reaction
generated through $a_0(980)/f_0(980)$ mixing.

\begin{acknowledgement}
  The authors wish to record their thanks to the COSY machine crew for
  producing the good experimental conditions needed for this experiment
  and also to other members of the ANKE collaboration for their help.  
  We are grateful to A.~Nogga for a careful reading of the manuscript.
  This work was supported by DAAD, DFG (436 RUS 113/630, 787, 940),
  RFFI (06-02-04013), the JCHP-FFE programme, and the HGF-VIQCD.
\end{acknowledgement}

%
%

\end{document}